\documentclass[baaa]{baaa}

\usepackage[pdftex]{hyperref}
\usepackage{subfigure}
\usepackage{natbib}
\usepackage{helvet,soul}
\usepackage[font=small]{caption}

\contriblanguage{1}

\contribtype{1}

\thematicarea{2}

\title{ 
A model for star formation in cosmological simulations \\ 
of galaxy formation 
}

\titlerunning{Star formation in simulations of galaxy formation}

\author{ 
E. Lozano\inst{1},  
C. Scannapieco\inst{1}  
\&  
S.E. Nuza\inst{2,1} 
}

\contact{lozano.ez@gmail.com}

\institute{
Departamento de Física, FCEN--UBA, Argentina
\and
Instituto de Astronomía y Física del Espacio, CONICET--UBA, Argentina
}

\resumen{En este trabajo presentamos un nuevo modelo para describir el proceso de formación estelar en galaxias, el cual incluye la descripción de las diferentes fases del gas -- molecular, atómica e ionizada -- en conjunto con su contenido metálico. Este modelo, que será acoplado a simulaciones cosmológicas de la formación de galaxias, será utilizado para investigar la relación entre la tasa de formación estelar (SFR) y la formación de hidrógeno molecular. El modelo sigue la evolución temporal de las tres fases en una nube de gas y estima la cantidad de masa estelar formada, a través de la resolución de un sistema de cinco ecuaciones diferenciales acopladas. Según lo esperado, encontramos una correlación fuerte y positiva entre la fracción molecular y la densidad de gas inicial, la cual se manifiesta en una correlación positiva entre la densidad de gas inicial y la SFR de la nube.}

\abstract{We present a new model to describe the star formation process in galaxies, which includes the description of the different gas phases -- molecular, atomic, and ionized -- together with its metal content. The model, which will be coupled to cosmological simulations of galaxy formation, will be used to investigate the relation between the star formation rate (SFR) and the formation of molecular hydrogen. The model follows the time evolution of the molecular, atomic and ionized phases in a gas cloud and estimates the amount of stellar mass formed, by solving a set of five coupled differential equations. As expected, we find a positive, strong correlation between the molecular fraction and the initial gas density, which manifests in a positive correlation between the initial gas density and the SFR of the cloud.} 


\begin{document}

\maketitle

\section{Introduction}
\label{sec:intro}

The star formation rate (SFR) is one of the most fundamental properties of galaxies. In a cosmological context, the SFR is the result of a complex network of processes that act together during a galaxy's lifetime, such as gas cooling, star formation, chemical enrichment and feedback from supernovae and galactic nuclei. Furthermore, the amount and properties of the gas from where stars form are strongly affected by mergers, interactions and mass accretion. Theoretical and observational studies show that the most important factor determining the star formation rate of a gas cloud is its density, although the details of this process are not yet fully understood.

Galaxy formation in the context of the standard cosmological paradigm, the $\Lambda$ Cold Dark Matter model ($\Lambda$CDM) is an extremely complex, highly non-linear process that is free from any simplifying symmetries. Numerical simulations are the methods best suited to study this process on a physical basis, since they can follow the joint evolution of dark matter and baryons, naturally capturing processes such as mergers and mass accretion. However, large uncertainties still exist in the treatment of the baryonic evolution, as the physical processes affecting this component -- star formation, various forms of feedback, chemical enrichment -- act at unresolved scales and are only introduced via sub-grid physics. These models then need to tune a number of free input parameters that are usually not independent from each other, making the predictions of different models sometimes inconsistent \citep{Scannapieco2012}.

Because of its central role in galaxy formation, it is extremely important for simulations to properly describe the star formation process at the resolved scales, together with the associated feedback and chemical enrichment processes, and investigate in more detail the assumed sub-grid models and their dependencies with the properties of the interstellar gas. In this work, we present a new model of star formation, which describes the star formation rate of a gas cloud considering its chemical abundance and the relative abundance of the atomic, molecular and ionized gas phases. This new model is designed to be coupled to our cosmological simulation code {\sc gadget3} \citep{Springel2008} and the extensions of \cite{Scannapieco2005,Scannapieco2006} and \cite{Poulhazan2018}. This model has already been applied to chemical evolution models -- simplified semi-analytic models of the formation of an isolated galaxy -- to study the properties of the Milky Way and chemical gradients \citep{Molla2017,Molla2018}.

This work is organized as follows. In Sec.~\ref{sec:model} we describe our new model; in Sec.~\ref{sec:results} we discuss the first results of the model and in Sec.~\ref{sec:conclusions} we present the conclusions.

\section{The star formation model}
\label{sec:model}

The new star formation model is based on the work of \cite{Ascasibar2015} (see also \cite{Molla2017,Murante2010,Millan-Irigoyen2020}), and is designed to follow the time evolution of various phases -- molecular, atomic and ionized -- in a gas cloud, also considering its metal content. In the context of our simulations, this model can be used to better describe the star formation process and investigate its efficiency as a function of the amount of gas in the molecular and atomic phases.
The model is intended as a replacement of the traditional prescription for the star formation rate used in galaxy formation simulations \citep{Katz1992}:

\begin{equation}
    \label{eq:eq1}
    \psi(t) = \epsilon_{\text{ff}}\,\frac{\rho}{t_{\text{ff}}} \, ,
\end{equation}
where $\epsilon_{\text{ff}}$ is the efficiency of star formation and $t_{\text{ff}}$ is the free fall time of the gas. In this way, $\psi(t)$ is only a function of the total gas density, $\rho$, and does not depend explicitly on the molecular or atomic fractions.

Our new model follows the coupled evolution of the different gas phases, ionized ($i$), atomic ($a$) and molecular ($m$) mass densities, together with the stars ($s$) and metals ($z$), solving a system of 5 coupled differential equations, namely:

\begin{align}
    & \frac{\text{d}i(t)}{\text{d}t} = - \frac{i(t)}{\tau_R} + (\eta_\text{ion} + R) \, \psi(t) \, ,                              \\
    & \frac{\text{d}a(t)}{\text{d}t} = - \frac{a(t)}{\tau_C} + \frac{i(t)}{\tau_R} + (\eta_\text{diss} - \eta_\text{ion}) \, \psi(t) \, , \\
    & \frac{\text{d}m(t)}{\text{d}t} = \frac{a(t)}{\tau_C} - (\eta_\text{diss} + 1) \, \psi(t) \, ,                                     \\
    & \frac{\text{d}s(t)}{\text{d}t} = (1 - R)\,\psi(t) \, ,                                                                         \\
    & \frac{\text{d}z(t)}{\text{d}t} = (Z_\text{SN}\,R - Z)\,\psi(t) \, .
\end{align}

\begin{figure}[!t]
    \centering
    \includegraphics[width=0.8\columnwidth]{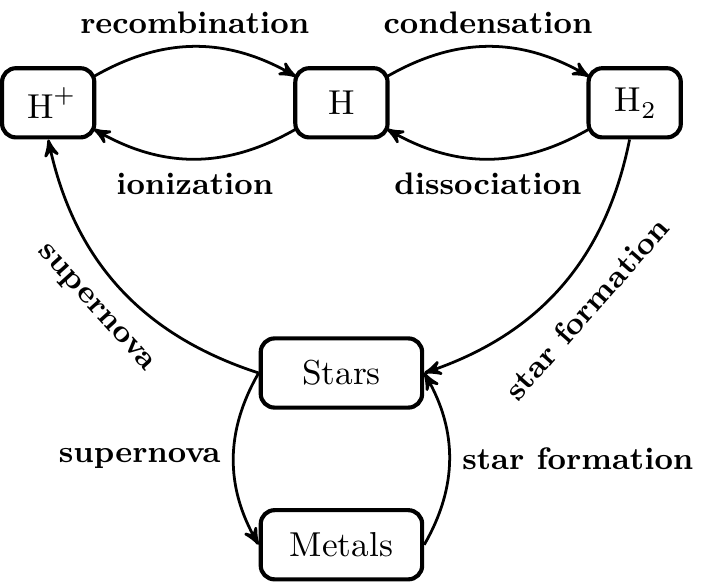}
    \caption{Physical processes relating the different phases. Arrows going inward indicate that the named process increases the amount of the corresponding phase, and vice versa for arrows going outward.}
    \label{fig:relations}
\end{figure}

These equations model the physical processes leading to a mass exchange between the different phases: photoionization of atomic hydrogen and supernovae -- with assumed efficiencies per unit of star formation rate $\eta_\text{ion}$ and $R$, respectively; recombination of ionized hydrogen with electrons -- characterized by a time $\tau_R$ --; dissociation of molecular hydrogen -- with an efficiency per unit of star formation rate $\eta_\text{diss}$ --; and condensation of atomic hydrogen -- catalyzed by dust grains \citep{Millan-Irigoyen2020} and regulated by the time parameter $\tau_C$. In our model, stars form from the molecular and metal components -- proportionally to $Z = z / (i + a + m)$, the fractional amount of metals --, with a typical timescale  $\tau_S$, and later feed metals and ionized gas to the interstellar medium (ISM) at their death as supernovae. This process is characterized by two parameters, $Z_\text{SN}$ for metal enrichment and $R$ for the ionized gas.
The processes leading to mass exchange between the different phases are schematized in Fig.~\ref{fig:relations}.

The model has several input parameters; however, most of them are constants that are well constrained empirically or theoretically that we take from the literature. There are two parameters, $\tau_R$ and $\tau_C$, which do depend on the properties of the gas, and this dependency is considered in our model following \cite{Osterbrock2006} and \cite{Millan-Irigoyen2020}.

By solving our system of coupled equations, we obtain $m(t)$, the amount of gas in the molecular phase as a function of time and, based on the strong observational correlation between molecular hydrogen and star formation \citep{Bigiel2008,Wong2002}, we assume a star formation rate of
\begin{equation}
    \psi(t) = \frac{m(t)}{\tau_S} \, ,
\end{equation}
for the gas cloud, which replaces the one given in Eq.~\ref{eq:eq1}. In this way, we have a new star formation law that couples the star formation rate and the molecular fraction.

\section{Results}
\label{sec:results}

As a first test of our model, we investigate the evolution of the different gas phases of a gas cloud in an idealized scenario. For this purpose, we integrated the system of equations for a single cloud of a given density, assuming initial parameters for the atomic, molecular and ionized gas fractions, and for the stellar and metal fractions. The system was evolved for a period of $1~\mathrm{Gyr}$ and for three different values of the initial gas density, $n = 1, \, 10, \, 100~\mathrm{cm^{-3}}$. We used, in particular, the following initial conditions: $i_f(t_0) = 0.001, \, a_f(t_0) = 0.998, \, m_f(t_0) = 0.001, \, s_f(t_0) = 0 \text{ and } z_f(t_0) = 0.0001$ where the subscript $f$ indicates the use of fractional quantities with respect to the total amount of gas, $g = i + a + m$. As mentioned above, $i$, $a$ and $m$ are respectively the mass densities of ionized, atomic and molecular gas.
Note that $g \equiv n$, given that $n$ is the number density of protons, so it can be readily transformed into mass density units.

\begin{figure}[!t]
    \centering
    \includegraphics[width=0.8\columnwidth]{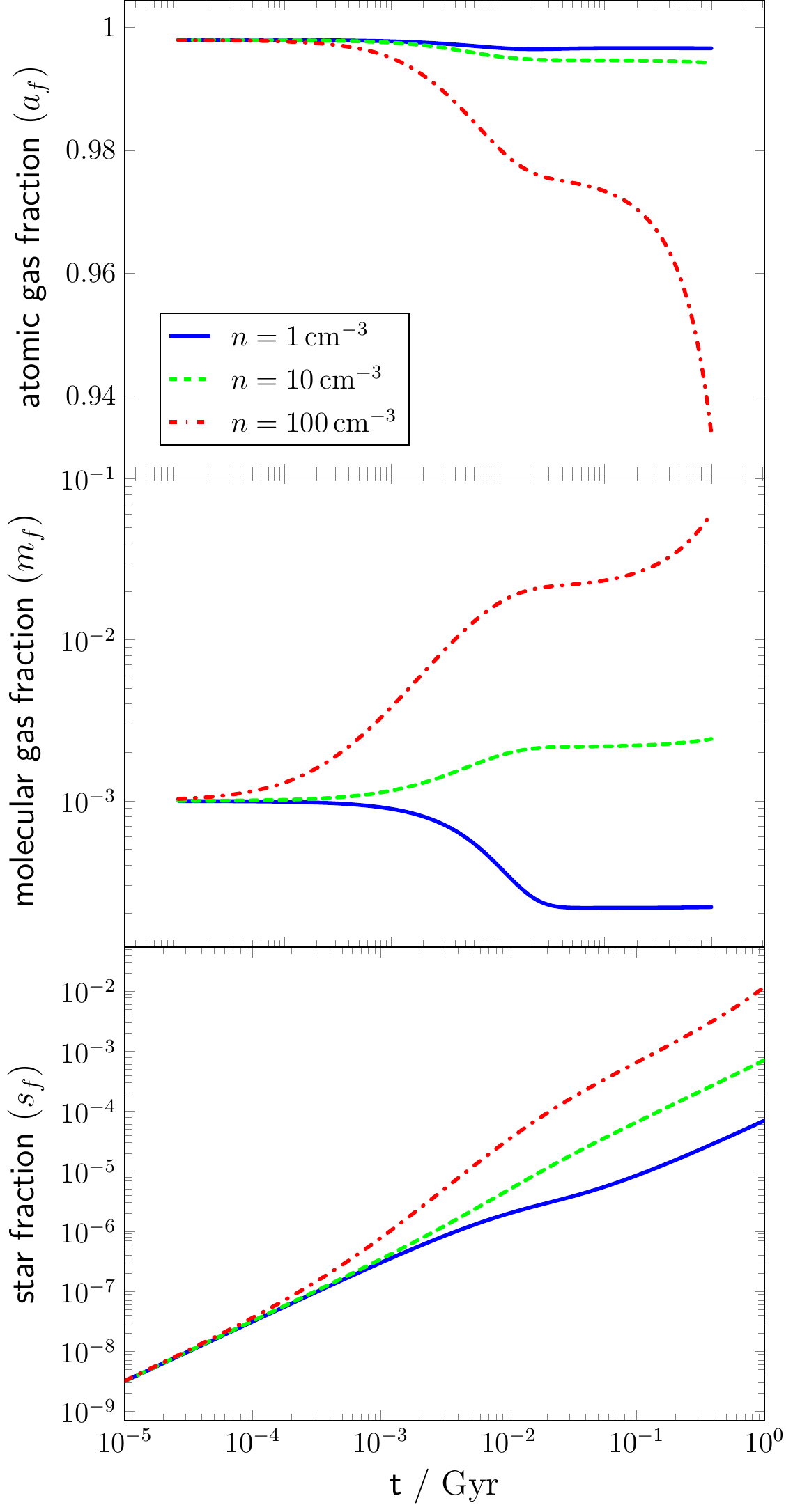}
    \caption{Evolution of atomic, molecular and stellar fractions, with respect to total gas, for three different values of initial total gas number density. \emph{Top panel:} Evolution of the atomic gas fraction. \emph{Middle panel:} Evolution of the molecular gas fraction. \emph{Bottom panel:} Evolution of the stellar fraction.}
    \label{fig:fractions}
\end{figure}

The results for the atomic, molecular and star fractions are shown in Fig.~\ref{fig:fractions}. We find a positive correlation between the initial density and the final fraction of molecular gas. For the highest density considered, the conversion of atomic to molecular gas is very efficient. For the other two densities, the atomic fraction is approximately constant, and the molecular phase cannot be replenished. These trends explain the behavior of the star fraction, which is higher for the highest density. After $1~\mathrm{Gyr}$ of evolution, the star fraction differs by an order of magnitude between each of the initial gas densities used.

We note that, although the amount of stellar mass formed from a gas cloud depends primarily on the total density, other initial values might also have a significant impact on the evolution of the molecular and stellar fractions. In particular, we have observed that the initial metallicity has a big impact on the ionized to molecular conversion efficiency, and consequently in the final star fraction. The higher the initial metallicity, the more stars form after a given period of time, maintaining the tendency already observed with the initial gas density.

\section{Conclusions}
\label{sec:conclusions}

We presented a new model to describe the star formation rate of a gas cloud, which considers the evolution of the atomic, molecular and ionized fraction of gas, providing an estimation of the stellar mass expected to form as a function of time. The model is specifically designed to be coupled to hydrodynamical simulations of galaxy formation in a cosmological context.

We tested our model on simplified scenarios, by applying it to the evolution of clouds with different initial properties. In particular, we studied the dependency of the molecular, atomic and stellar fractions of a gas cloud, composed at the initial time almost entirely by atomic material, assuming three different values for the total initial density. Our results showed a strong dependence of the amount of stellar mass formed in the cloud and the initial density, which follows the trends found for the atomic and molecular phases.

Our model has already been applied in simulations of the formation of a Milky Way-mass galaxy -- both in an idealized scenario and in cosmological context. Compared to the standard model (Eq.~\ref{eq:eq1}), we find that coupling the star formation efficiency to the amount of molecular gas produces galaxies with lower stellar masses and with retarded star formation rates (Lozano et al., in preparation).

\begin{acknowledgement}
    The authors acknowledge support by the Agencia Nacional Científica y Tecnológica (ANPyCT), PICT-201-0667.
\end{acknowledgement}


\bibliographystyle{baaa}
\small
\bibliography{bibliography}

\end{document}